\documentclass[sigplan,10pt]{acmart}

\acmJournal{PACMPL}
\acmVolume{1}
\acmNumber{CONF} 
\acmArticle{1}
\acmYear{2018}
\acmMonth{1}
\acmDOI{} 
\startPage{1}

\acmConference{Onward! Essays 2020}{October 2020}{Chicago, IL, USA}

\setcopyright{none}

\usepackage{booktabs}   
\usepackage{subcaption} 
\usepackage{todonotes}
\usepackage{epigraph}
\usepackage{dirtree}
\usepackage[normalem]{ulem}
\usepackage{tikz}
\usepackage{float}

\begin{document}
	
	\title{Putting the Semantics into Semantic Versioning}         

	
	\author{Patrick Lam}
	\affiliation{
		\institution{University of Waterloo}            
	}
	\email{patrick.lam@uwaterloo.ca}          
	
	\author{Jens Dietrich}
	\affiliation{
		\institution{Victoria University of Wellington}           
	}
	\email{jens.dietrich@vuw.ac.nz}         
	
	\author{David J. Pearce}
	\affiliation{
		\institution{Victoria University of Wellington}           
	}
	\email{djp@ecs.vuw.ac.nz}         
	
	\begin{abstract}
          The long-standing aspiration for software reuse has made astonishing strides in the past few years. Many modern software development ecosystems now come with rich sets of publicly-available components contributed by the community. Downstream developers can leverage these upstream components, boosting their productivity.

However, components evolve at their own pace. This imposes obligations on and yields benefits for downstream developers, especially since changes can be breaking, requiring additional downstream work to adapt to. Upgrading too late leaves downstream vulnerable to security issues and missing out on useful improvements; upgrading too early results in excess work. Semantic versioning has been proposed as an elegant mechanism to communicate levels of compatibility, enabling downstream developers to automate dependency upgrades. 

While it is questionable whether a version number can adequately characterize version compatibility in general, we argue that developers would greatly benefit from tools such as semantic version calculators to help them upgrade safely. The time is now for the research community to develop such tools: large component ecosystems exist and are accessible, component interactions have become observable through automated builds, and recent advances in program analysis make the development of relevant tools feasible. In particular, contracts (both traditional and lightweight) are a promising input to semantic versioning calculators, which can suggest whether an upgrade is likely to be safe.




	\end{abstract}



	\maketitle
	
	\section{Introduction}

\epigraph{Every change is an incompatible change. A risk/benefit analysis is always required.}{Martin Buchholz\footnote{https://blogs.oracle.com/darcy/kinds-of-compatibility:-source,-binary,-and-behavioral}}

Modern software is built to depend on countless subcomponents, each with their own lifecycles. \citeN{cox2019surviving} writes about surviving software dependencies from a practitioner's point of view, and provides advice about how to evaluate, incorporate, and maintain one's dependencies. 
As Cox writes, in today's world, responsible developers must upgrade their dependencies in a timely fashion. Upgrades create a conundrum for developers: they pitch agility against predictability. Developers appreciate automatic upgrades which deploy improved versions of their upstream dependencies. Developers do not appreciate upgrades which introduce unexpected consequences and new faults. Tools can help developers accept safe upgrades with confidence and focus attention on potentially-problematic upgrades.

\begin{figure}[H]
	\includegraphics[width=0.7\columnwidth]{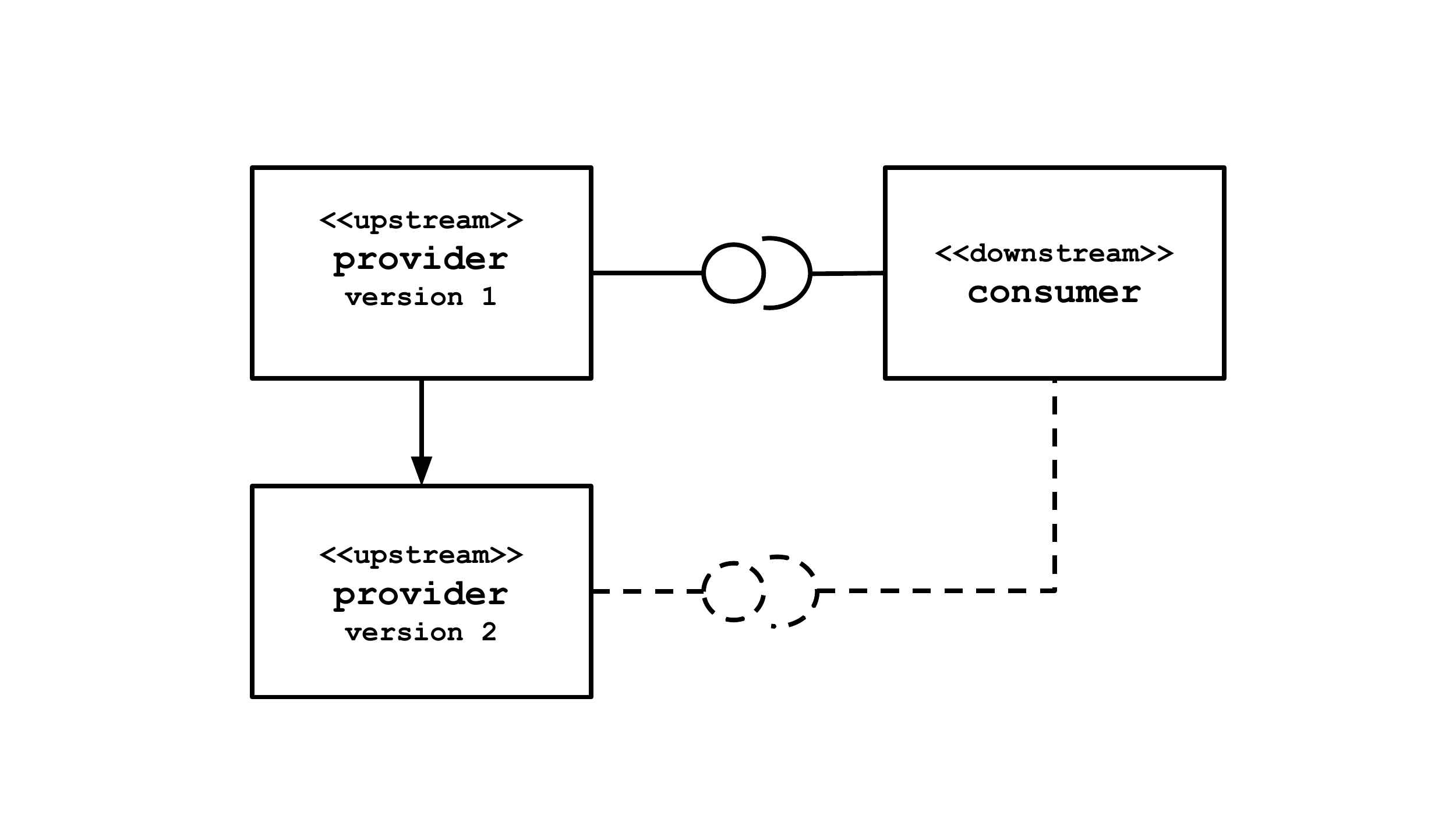}
	\caption{Upgrading from version 1 to version 2 of provider. Both upstream and downstream wish to avoid breakage.}
	\label{fig:overview}
\end{figure}

Figure~\ref{fig:overview} depicts the overall process of dependency evolution. For lack of a better alternative, we base this on a UML component diagram. The vertical arrow represents evolution. In this scenario, a consumer component has a dependency on an upstream component that is working (i.e., it has been successfully deployed, satisfying some kind of explicit or implicit contract). When the provider component is changed and deployed as version 2, its developer must make a judgment about whether this change is compatible, in the sense that it will not break \textit{any} client (consumer) downstream. In the case of an incompatible change, the upstream developer must ensure that the nature of the change, in particular the level of breaking changes, is clearly communicated to downstream developers.  Next, downstream developers faced with any change also have a task---they must decide whether that change is compatible with \textit{this particular client}. The downstream developer could trust the upstream developer and forego this step. However, often a downstream developer has learned from painful experience that the checks of the upstream developer are not sufficient and cannot be trusted. An experienced downstream developer will therefore run additional checks, for instance, by running integration tests.

This essay argues that the research community has an important role to play in this increasingly relevant aspect of modern software development, by providing a suite of tools that help both upstream and downstream developers with evolution-related tasks. We use a generalized meaning of ``tool'' which includes not just a concrete program that one executes on one's software but also abstract methods and concepts such as contracts and versioning schemes.

Indeed, versioning schemes are an important technique that is already widely used, especially in the form of semantic versioning~\cite{preston2013semantic}. Declarative dependency declaration and semantic versioning promise an elegant solution to the software upgrade conundrum: upstream component providers use version numbers to encode compatibility levels of the respective components, while downstream component consumers declare dependencies using ranges of compatible versions that ensure automated upgrades do not introduce new faults.

One can think of semantic versioning as a condensed code contract. A safe upgrade is one for which contracts do not change incompatibly. The complexity of component interactions in modern systems suggests that computing version numbers which clients can rely on is challenging-to-impossible. However, tools can help improve the current practice, even if imperfect.

As a partial aid to versioning, we advocate for the development of \textit{semantic version calculators} (or \textit{version calculators} for short). Such a version calculator must, given a software artifact $A$ with version number x.y.z, and a different version $A'$ of $A$, compute a new version number x'.y'.z', which accurately communicates the changes made to X according to the rules of semantic versioning.

The implication of semantic versioning is that clients may rely on dependencies subject to flexible version constraints, like 1.2.* (``any version where the major version is 1 and the minor version is 2''). Such a client may safely upgrade to new micro versions (e.g., from 1.2.3 to 1.2.4). The actual upgrading can thus be fully automated and performed by a dependency manager, usually at build-time\footnote{It is also common to perform ``hot'' component upgrades where systems must run in 24/7 mode. \citeN{hicks01:_dynam_softw_updat} proposed a general-purpose methodology for dynamic updates, while in practice, OSGi supports (less-principled) dynamic updates based on the use of class loaders swapping class definitions.}. There exist services which carry out such updates today; at the time of writing, Snyk\footnote{\url{https://snyk.io/}} provided this service for npm code. Assuming that the quality of components generally increases as they evolve, this keeps the deployment process of programs agile (highly automated) while program behaviour remains predictable.

Assuming that version numbers tell the whole story (Section~\ref{sec:version-numbers} discusses relationships between version numbers and software evolution in more detail), the challenge is then to compute version numbers. If we assume that all that clients care about is shallow contracts that operate on the level of API signatures in a statically typed language, then providing tool support is well within the state of the art. This tool-based approach has enjoyed some popularity among developers. For instance, Java developers have already adopted existing tools like \textit{clirr}\footnote{\url{http://clirr.sourceforge.net/}} and \textit{Revapi}\footnote{\url{https://revapi.org/}}. These tools provide value by helping to avoid errors caused by violating the rules of binary compatibility~\cite[Sect.5.4]{lindholm14:_java_virtual_machin_specif_java_se_edition}, which many developers struggle to understand~\cite{dietrich2016java}. Our pilot study in Section~\ref{sec:upstream} observes linking issues caused in practice which could have been prevented with such tools.

Unfortunately, shallow contracts are only the tip of the iceberg. More complex component contracts matter a lot. Examples include issues related to code behaviour, quality of service attributes, and more~\cite{beugnard1999making}. As an example, consider the contract implied by the licenses of a program and the components it uses.  Making the license of a component less permissive (e.g. from MIT to GPL) would certainly be considered a breaking change that would warrant a major version upgrade. Making such a change in a minor version would be considered as highly undesirable or even hostile by clients, primarily because tools are empowered to automatically update dependencies across minor version changes. Clients would not want their software to become undistributable without warning.

A more common scenario is when the behaviour of components changes in a way that affects the behaviour of their clients.  Examples are strengthened precondition checks leading to exceptions, changes in the traversal order to collection data structures returned to clients, or dropped null pointer safety in methods returning values. A good way to conceptually capture many of these scenarios is to describe them in terms of \textit{safe substitution}, similar to Liskov's Substitution Principle for behavioural subtyping~\cite{liskov1994behavioral}, but instead applied to the safe substitution of an artifact (class, method, procedure, function, etc) by a later version of itself.


Detecting changes that predict the impact on potential clients and calculating a version number to communicate the impact is a program analysis problem. As with any program analysis problem, a perfect solution is not feasible~\cite{rice1953classes}, and researchers must develop practical approximations that achieve acceptable precision and soundness trade-offs.

In the context of semantic version calculation, we recast precision and soundness as \textit{optimistic} and \textit{pessimistic calculations}, respectively. An optimistic calculation is based on an unsound analysis---it may miss changes which could break clients, and some clients could upgrade too aggressively, potentially requiring developers to rollback or mitigate changes.  On the other hand, a pessimistic calculation is based on an imprecise analysis---it would, for instance, increase the major number of a version often, even if the respective changes in the component have no negative impact on most clients. This pushes the burden onto downstream developers, who are supposed to manually investigate major number changes for incompatibility. Excessive false positives will eventually lead developers to ignore legitimate potentially breaking changes. To put it another way, an overly-optimistic calculator has developers cleaning up issues that were not flagged, while an overly-pessimistic calculator results in developers needlessly inspecting changes that are harmless in practice.

In general, program analysis aims for a good balance between soundness (or recall, if used as a quantitative term) and precision. Many analyses can tolerate both unsoundness and imprecision. An example is vulnerability detection: as long as some (but not necessarily all) vulnerabilities are found, and the ratio between false alerts and detected vulnerabilities is acceptable, the analysis is generally considered useful. In the context of semantic version calculation, the situation is even more complex. Different strategies influence how quickly fixed faults in libraries are propagated to clients (faster in the optimistic, slower in the pessimistic approach); and how likely compatibility-related faults are to occur in the program (more likely in the optimistic, and less likely in the pessimistic approach). The consequences affect the choice of strategy in the design of calculation tools.

\paragraph{Thesis.}   
We argue that, given that
developers in today's software ecosystem constantly need to evaluate
version compatibility, \emph{the research community has a key role to
  play in developing relevant tools to help developers}. From the
ecosystem side, the availability of releases and tests in
repositories, continuous integration, and issue trackers; and from the
technology side, recent advances in powerful yet scalable program
analysis, both combine to now enable the development of practical
tools to manage versioning. These tools can address pain points in
modern software development and have a significant impact on practice.


	\section{Status Quo: Three Challenges for API~Upgrading}
\label{s:status_quo}

Let's have a closer look at the fundamental conflict in upgrade timing that developers face. We start with the component (upstream) developers' perspective in Section~\ref{sec:upstream}, exploring their role in (inadvertently) breaking clients. Shifting to the (downstream) client perspective, upgrading late (Section~\ref{sec:status-quo:tooslow}) will result in unpatched vulnerabilities and bugs. On the other hand, upgrading early (Section~\ref{sec:status-quo:toofast}) requires client developers to act earlier than they otherwise would, either in the form of dependency inspection or adaptation to new APIs. Developers of clients must balance optimism and pessimism, and it is our thesis that they would greatly benefit from the availability of tools to help them do so.

\paragraph{About components.} We often refer to ``components'' in this essay.
Our intended meaning is broad, and we include artifacts at all
levels, from the kernel/application interfaces, to dynamically-linked
libraries loaded through the system dynamic linker, up to higher-level
libraries as in Java modules and npm packages consisting of code and metadata, 
all the way to components providing REST APIs
which enable distributed systems to communicate. Similar considerations apply
to languages, and we discuss them briefly in Section~\ref{sec:upstream}, but our focus
is on components.

We focus particularly on components that
exist in discoverable ecosystems, such that a client can select a
desired version of its dependency, either manually, or automatically by employing some dependency resolution mechanism.

For our purposes, component interfaces are documented in published
API specifications.  These specifications are contracts (both formal
and informal); we further discuss contracts in
Section~\ref{sec:about-contracts}. The most important aspect to these
components for the purpose of this essay is that component developers
periodically release labelled versions of their components, each
version having its own API.

\paragraph{About semantic versioning.} \citeN{preston2013semantic} defines semantic versioning
as a system where components' version numbers
have three parts: major.minor.patch. The major version indicates a breaking change; the semantic versioning specification calls for
the major version to be increased ``if any backwards incompatible changes are introduced to the public API''.
The minor version is to be increased when new, backwards-compatible, public functionality is added or when some
public API is deprecated; it may also denote ``substantial new functionality or improvements''. The patch version
is to be increased if ``only backwards compatible bug fixes are introduced'', where a bug fix is ``an internal change that fixes incorrect behaviour''.

Semantic versioning is concerned with changes in an upstream component that affect any possible (API-respecting) downstream client.
An incompatible change must be marked as a major version change.
A compatible change---where new version $A'$ still conforms to the contract of old version $A$---may be
a minor or patch upgrade. If the contract of $A'$ is stronger than that for $A$, then $A'$ is a minor
upgrade. If the contracts are identical, then $A'$ is a patch upgrade.

While the semantic versioning specification is written in black-and-white terms, we will describe (in this section) some
inadvertent breaking minor changes and then subsequent minor changes that revert the breaking changes. Technically this breaks the
semantic versioning specification. In practice, we find developers tend to reserve major changes for planned significant breakages.



The version compatibility problem at the core of this essay reduces to the following question. For software artifact $A$, does version $A'$ introduce any changes that would break any possible client of $A$? Semantic versioning is a tool for upstream developers to communicate with downstream developers about the impacts of the changes they have made.

\subsection{Why and How Upstream Breaks APIs}
\label{sec:upstream}


We first consider component evolution from the upstream perspective. Upstream developers are ostensibly responsible for doing no harm, on balance. Of course, some upgrades fix bugs. Yet, some upgrades break clients.

\paragraph{Why?} Sometimes, developers are unaware that they are making breaking changes. For instance, \citeN{dietrich2016java} surveyed 414 Java developers and found that these developers have only a limited understanding of Java's binary compatibility rules. This suggests that some breaking changes are inadvertent and could be mitigated by better tool support.

Other changes are deliberate.
Reusable components, like all programs, are subject to change pressure. As \citeN{lehman1980programs}'s law of continuing change puts it: ``A program that is used and that as an implementation of its specification reflects some other reality, undergoes continual change or becomes progressively less useful.'' \citeN{dig06:_how_apis} manually analyzed breaking API changes in four Java components and found that many breaking changes were refactorings. More recently, \citeN{brito2018and} qualitatively studied the reasons that developers made deliberate API-breaking changes, and found that the main reasons were the implementation of new features, the simplification of APIs, and the improvement of maintainability. Occasionally, changes are malicious, as in the case of an upgrade to the npm event-stream library which aimed to steal cryptocurrency from wallets maintained in a particular client~\cite{wayne19:_stamp}.

There is a grey area between deliberate and inadvertent breaking changes. Sometimes, a particular change may be allowed according to the published API, and hence should not be breaking. Yet dependencies may rely on implementation-specific behaviour that is undocumented or contrary to documentation; we will discuss an example involving Firefox and its upstream library fontconfig later in this subsection.


\paragraph{How?} Communities have different conventions about how to make API-breaking changes. Developer behaviour is heavily influenced by social factors like policies and practices. \citeN{bogart2016break} surveyed three communities and found that: (1) in Eclipse, developers try not to break APIs, i.e. there is a strong focus on API stability; (2) in R/CRAN, developers notify downstream developers; (3) and in Node.js/npm, developers increase the major version number when making breaking changes.

Considerations about breaking changes also apply at a language level. Breaking changes do occur; for instance, F\# documented two breaking changes going from version 4.1 to 4.5\footnote{\url{https://devblogs.microsoft.com/dotnet/announcing-f-4-5/}}. More generally, language design committees publish statements about the conditions under which they are willing to entertain breaking changes. ISO/IEC JTC1/SC22/WG14, responsible for C, published a charter for C2x which aims to ``avoid {`quiet changes'}'' and to ``minimize incompatibilities with C90''.
Some members of the C++ working group recently published a proposal\footnote{\url{http://www.open-std.org/jtc1/sc22/wg21/docs/papers/2020/p2137r0.html}} which states that their vision of C++ includes ease-of-migration rather than compatibility between versions as goals for the C++ language itself. Other members have proposed an ``epoch'' mechanism for C++\footnote{\url{http://www.open-std.org/jtc1/sc22/wg21/docs/papers/2020/p1881r1.html}} to preserve backwards compatibility while enabling evolution.


This suggests that, to some extent, upstream developers in some communities do attempt to limit or mitigate API-breaking (and language-breaking) changes. In Sections~\ref{sec:about-contracts} and~\ref{sec:version-numbers}, we discuss extant tools and suggest novel tools to meet this important need.

\paragraph{Example: A semi-inadvertent breaking change in C.} Firefox uses the fontconfig library to select fonts. A commit to fontconfig 2.10.92 caused it to reject empty filenames, thus breaking Firefox's font display\footnote{\url{https://bugzilla.mozilla.org/show_bug.cgi?id=857922}}. The API documentation (plain English text in Doxygen format) was silent as to whether empty filenames were allowed or not. Once this client behaviour was observed, the upstream library amended its API so that empty filenames were explicitly allowed and modified its implementation accordingly.


\paragraph{Survey: Breaking platform-level changes in .NET.}
For the .NET platform, \citeN{msft:_break_chang_defin,msft:_break_chang_rules} has defined the notion of a breaking change. It publishes a detailed list of breaking changes with each new platform release. We summarize here the 77 documented breaking changes between .NET 2.2 and 3.0\footnote{\url{https://docs.microsoft.com/en-us/dotnet/core/compatibility/2.2-3.0}} to better understand the types of deliberate breaking changes that occur in practice; after each category name, N denotes the number of breaking changes in that category. Some changes belong to multiple categories.

\begin{itemize}
\item Behaviour change (N=31): the behaviour of the component changes in a potentially-breaking way, or the client is now required to carry out some additional actions.
\item API replacement/redirection (N=30): the component would like its clients to use a new API in place of a previously-published API; sometimes done automatically with .NET type forwarding.
\item Published API removal (N=19): the component retracts a given previously-published public API; in some cases, no replacement API is provided.
\item Reflecting external world change (N=6): some external component changed (e.g. Bootstrap 3 to 4) and .NET made this change visible to clients.
\item Unpublished API removal (N=5): ASP.NET previously contained public APIs in {\tt .Internal} namespaces, removed in .NET 3.0.
\item Remove flexibility (N=4): component publishers decided to e.g. make classes sealed, that is, no longer extensible in subclasses.
\end{itemize}

\paragraph{Examples: Upstream breaking changes in Java.}

It is not at all hard to
find examples in the wild where breaking changes have been
inadvertently introduced between minor or micro versions.
We illustrate some Java examples.

JSoup v1.10.1 included a performance refactoring for
``\emph{reducing memory allocation and garbage collection}'' which
introduced a breaking change over the previous version
(v1.9.2)\footnote{\url{https://github.com/jhy/jsoup/issues/830}
}. Likewise, v3.3 of the Apache Commons Math library
introduced a fix for floating point
equality.\footnote{\url{https://issues.apache.org/jira/browse/MATH-1118}}
Unfortunately, the fix was botched and not applied consistently across
the code base, leading to downstream
failures.\footnote{\url{https://issues.apache.org/jira/browse/MATH-1127}}
Another fix for date handling in Apache Commons Lang
introduced between v3.2.1 and v3.3 again led to problems
downstream.\footnote{\url{https://issues.apache.org/jira/browse/LANG-951}}
This time, while the fix was done correctly, it nevertheless impacted
downstream clients unexpectedly.  Finally, a fix between v2.6.0
and v2.6.1 of the FasterXML library for JSON
processing introduced breaking changes which were quickly spotted by
multiple downstream
clients.\footnote{\url{https://github.com/FasterXML/jackson-module-scala/issues/222}}\footnote{\url{https://github.com/FasterXML/jackson-databind/issues/954}}\footnote{We discovered those examples using cross-version testing, a technique discussed in Section~\ref{sec:formalcontracts}}.




Another concrete example can be found in asm, a popular bytecode engineering and analysis library~\cite{bruneton2002asm}.  In commit 
38097600\footnote{see \url{https://gitlab.ow2.org/asm/asm/-/commit/<id>}}, all packages were renamed, replacing the ``objectweb'' token by  ``ow2'', thus requiring compensating changes in all clients wishing to upgrade.  A more subtle change occurred in commit 7bc1be02, when the  \texttt{ClassVisitor} type widely used by clients was changed from an interface to an abstract class. This caused numerous issues in clients encountering \texttt{IncompatibleClassChangeError} when upgrading from asm-3.* to asm-4.*, for instance \url{https://github.com/bmc/javautil/issues/17} . These changes were then released in the next versions following these commits, and started to affect clients.

Another interesting change occurred in commit 291f39aa. Here, the license was made more permissive---from LGPL to BSD---and the version number went from 1.3.5 to 1.3.6. 

\paragraph{Pilot study: incidence of breaking changes in Java.} In the presence of breaking changes, upgrading requires developers to adapt their downstream components to changes in upstream.
To estimate how much adaptation developers need to do, we conducted a simple experiment. Since incompatibilities often result in specific errors or exceptions that usually occur at component boundaries, searching GitHub issues offers some insights into the scale of the problem. The GitHub search API supports issue subsystem queries, returning a count of the matching issues for search strings, amongst other information. For instance, to search for issues mentioning \texttt{java.lang.NoSuchMethodError}, an exception thrown by the JVM when no matching methods are found and usually caused by a change in a method signature between versions, the following query can be used:
\begin{center}
  {\scriptsize \texttt{https://github.com/search?q=java.lang.NoSuchMethodError\&type=Issues} .}
\end{center}

\begin{table}[]
	\begin{tabular}{lr}
		error or exception                          	& 	issues   \\ \hline
		java.lang.NullPointerException     		& 	199,334 \\
		java.lang.ClassCastException       		& 	81,333  \\
		java.lang.IllegalArgumentException	& 	79,427  \\
		java.lang.IllegalStateException 			& 61,031 \\
		\textbf{java.lang.NoSuchMethodError	}		  	&		\textbf{30,450} \\
		java.lang.OutOfMemoryError			   & 	24,711 \\
		java.lang.StackOverflowError 			 & 	11,549 \\
		\textbf{java.lang.NoSuchFieldError}				& \textbf{3,872} \\ 
		\textbf{java.lang.IncompatibleClassChangeError}	& \textbf{2,962} \\
	\end{tabular}
	\caption{Number of GitHub issues referencing common Java exceptions and errors, ordered by issue count  (queries performed on 16 March 2020).}
	\label{table:issues}
\end{table}

Table~\ref{table:issues} shows the number of issues referencing some common Java exceptions and errors. Admittedly, this number is without reference to a denominator; however, it still shows that many users and developers are affected by these issues. Bolded rows represent potential contract violations between client and library code. We include some common exceptions and errors for comparison and perspective. \texttt{\texttt{NoSuchMethod-}} / \texttt{NoSuchField-} and \texttt{IncompatibleClassChangeError} are errors thrown by the JVM to signal binary compatibility violations, so those errors almost certainly occur at component boundaries due to upgrades---the compiler would prevent such situations within components. The situation is less clear for \texttt{Illegal\-Argument-} and \texttt{IllegalStateException}. Those exceptions are normally used to signal precondition violations, which are likely to occur at component boundaries between clients and libraries, assuming that intra-component occurrences are detected during regression testing. 

Similarly, searching for issues with GitHub's internal engine using the search phrase ``error after upgrade''\footnote{\url{https://github.com/search?q=error+after+upgrade&type=Issues}, query performed on 16 May 2020} yields 2,669,047 issues, and sampling the results suggests that the query results have a high precision. These numbers hint that issues that occur at component boundaries are significant pain points for developers.

\paragraph{Community practices to avoid breaking changes.}
Practices are shaped, in part, by the affordances of the language.
We have seen how, in its role as an upstream developer, \citeN{msft:_break_chang} explicitly documents breaking changes between versions of its released .NET APIs and suggests that library developers do the same. .NET changes are manually tagged and documented by upstream developers and collected into a document that is published with each release. It is currently the responsibility of the upstream developers to be aware of when their changes may potentially be breaking; our thesis is that tool support can improve this process. 
An earlier version of OpenJDK documentation\footnote{\url{http://cr.openjdk.java.net/~darcy/OpenJdkDevGuide/OpenJdkDevelopersGuide.v0.777.html}} identified three kinds of Java SE releases: ``platform'', ``maintenance'', and ``update'', and the criteria for changes to be included in each of these releases. Platform releases include major, breaking changes; behavioural changes should only occur in platform releases. Maintenance and update releases implement the same Java specification but contain bug fixes, more in a maintenance release than an update release. Hadoop includes annotations to indicate whether an API is stable, evolving (may change in minor releases), or unstable, along with the intended audience for an API (somewhat paralleling Java 9 modules, which we discuss in Section~\ref{sec:api-surfaces}). JUnit 5 labels its APIs as internal, deprecated, experimental, maintained (will not change in at least the next minor release), or stable (will not change in the current major release).

At a language level, Java checks for consistency between client and library code at both compile and link time. The idea is to catch certain potential API-breaking changes before runtime and to signal them as compiler or linker errors~\cite[sect 13]{jls8spec}. Java's checks primarily flag changes based on the types associated with methods and fields, along with a few other properties such as visibility; behavioural properties remain the responsibility of both component and client developers. That is no different from the fontconfig example above, which was in the C context. Several studies have looked into API syntax changes between Java library versions that can break clients~\cite{dietrich2014broken}, \cite{raemaekers2014semantic}.

The Java linker misses subtle upstream signature changes which may affect downstream runtime behaviour. Examples include changes to type parameters resulting in class cast exceptions; addition of checked exceptions to method signatures\footnote{Neither type parameters nor checked exceptions are checked during linking.}; changes to constants which are inlined by the compiler; and more complex cases like stack overflows caused by runtime resolution of overriding methods with covariant return types~\cite{robertson2013hazards}, \cite{jezek2016magic}.

Many of the remaining breaking signature changes can be detected using a relatively simple static analysis operating on type signatures.  \textit{Revapi}\footnote{\url{https://revapi.org/}} is one such example which can be embedded into a (e.g. \textit{Maven}) build script.  This tool looks for various changes between two versions of a jar file, including the removal of public classes/methods/fields, or changes to the signatures of public methods/fields, etc.  Indeed, the API syntax change studies mentioned above used such an approach and found many \textit{potentially} breaking changes that may affect some clients.  Furthermore, such tools are gaining more widespread uptake in industry (e.g. by Palantir~\cite{r.19:_preven_java_api_abi}).


\subsection{The Perils of Upgrading Late}
\label{sec:status-quo:tooslow}

We now shift perspective from upstream component developers to downstream client developers, who must decide when to upgrade. Upgrading late leaves clients exposed to their dependencies' bugs and vulnerabilities, and unable to profit from the dependencies' routine enhancements. A further disadvantage of upgrading late is that the client developers are less in control of when they upgrade---while a security patch is always time-critical, a client must first upgrade to a supported version that accepts the security patch, and that first upgrade can be done proactively. Stale dependencies are one of the greatest risk factors in application security: the OWASP Top Ten list includes ``A9. Using Components with Known Vulnerabilities''~\cite{owaspTop102017}.


The use of outdated dependencies is often described as \textit{technical lag}---the time period, or ``version distance'', between the release of the version of a component it depends on, and that of the latest version. Technical lag can be considered to be a form of technical debt that accrues with the passage of time, as the world changes around the client. It is an indicator of the resistance of developers to upgrade, and reflects on the (perceived) likelihood that an otherwise beneficial upgrade will break a system due to an incompatibility that would require intervention.  Technical lag has been identified as a major source of security vulnerabilities. For instance, a study on client-side use of JavaScript found that use of outdated libraries is common, and 37\% of web sites (from a data set of 133k sites) used at least one library with a known vulnerability. \citeN{cox2015measuring} measured that projects using outdated dependencies were four times more likely to have security issues than those with up-to-date dependencies. \citeN{pashchenko2018vulnerable} noted that the vast majority (81\%) of vulnerable dependencies may be fixed by simply upgrading to a new version.
\citeN{duan17:_ident_open_sourc_licen_violat} found that out of 1.6 million free Google Play Store apps, over 100,000 of them used known vulnerable versions of libraries, despite Google's App Security Improvement Program (ASIP) requiring developers to upgrade their dependencies (by banning future app updates). They did, however, find that ASIP did appear to reduce usage of vulnerable library versions.

As a concrete example, around 2015, numerous vulnerabilities were discovered in applications using Java binary serialization. These vulnerabilities used gadgets to pass data from object streams via proxies and reflection widely used in libraries like Apache commons collections to unsafe sinks like \texttt{Runtime::exec}, enabling arbitrary code execution attacks. This affected major web frameworks and application servers, including jenkins (CVE-2017-1000353) and weblogic (CVE-2015-4852). 

The social and economic consequences of upgrading late can be dire. In 2017, a vulnerability in the Apache struts web framework (CVE-2017-5638) was exploited in an attack on Equifax. Quoting from the testimony of Richard F. Smith, the CEO at the time, before the U.S. House Committee on Energy and Commerce Subcommittee on Digital Commerce and Consumer Protection~\cite{smith2017testimony}:

\begin{quote}
	``\emph{We now know that criminals executed a major cyberattack on Equifax, hacked into our data, and were able to access information for over 140 million American consumers. The
	information accessed includes names, Social Security numbers, birth dates, addresses, and in some instances, driver's license numbers; credit card information for approximately 209,000 consumers was also stolen, as well as certain dispute documents with personally identifying information for approximately 182,000 consumers.}''
\end{quote}

Incidents are always caused by a chain of mishaps. It seems that, in this case, the chain included the relevant machines not being scanned in a network scan, struts not being detected as being present on the machines, and notifications about the vulnerability going to the wrong employees. The result was that Equifax failed to upgrade the dependency to the version with the patch~\cite{office18:_gao_data_protec}.



\subsection{The Perils of Upgrading Early}
\label{sec:status-quo:toofast}

Historically, from an end-user point of view, it is well-known that .0 releases are high-risk; prudent users do not live at the bleeding edge, but rather wait for inevitable patch releases.

So, why do clients not constantly upgrade even if these upgrades can be automated? No change is risk-free. Changes introduced by upgrades (whether major or minor) can and do introduce compatibility bugs\footnote{An industry contact estimates that, in his experience, about 1 in 5 library upgrades are problematic.}. As mentioned above, deferring changes results in accruing technical debt. Eagerly upgrading dependencies can therefore be considered as an attempt to proactively pay off technical debt. It is not free and often requires developer intervention to work around the introduced compatibility bugs. Reasons to not proactively upgrade might be as follows:
\begin{itemize}
  \item the upgrade might not provide any immediate benefit to the client (paying technical debt diverts engineering effort from other tasks that may be more immediately needed---in other words, technical debt doesn't matter if one is about to declare bankruptcy).
\item the client developers may have inadequate continuous integration/continuous delivery/rollback infrastructure, making them unable to discover problems before releasing to production (so each upgrade is risky);
  \item the upgrade process itself may require non-trivial recompilation, as documented for React Native\footnote{\url{https://github.com/react-native-community/discussions-and-proposals/issues/68
  }} (so each upgrade is slow);
  \item the upgrade might uncover an upstream bug that the upstream developers eventually fix on their own (i.e. some problems fix themselves, from the client perspective; the Firefox/fontconfig bug was one such example, if one was a non-Firefox client of fontconfig); and,
  \item common upgrade problems are better documented on the Internet if one is not the first to encounter them (i.e. late upgrades are easier for programming-by-Web-search).
\end{itemize}

\subsection{Automating upgrades}

From the library client point of view, developers of client code would generally prefer to do less dependency management. To that end, many package managers support some form of automated upgrade of dependencies. Developers supply dependency constraints, and package managers resolve the constraints to actual versions.  If the constraints happen to match multiple versions available in a repository, then the package manager can apply heuristics to choose the \textit{best version }available satisfying the constraint. Examples of dependency constraints include version ranges as well as wild cards for the minor or micro parts in major.minor.micro-style versioning schemes. There are two use cases for underconstrained version specifications: (1) to facilitate conflict resolution for multiple, potentially conflicting versions of the same dependency\footnote{Java developers often refer to the problems arising from this as \textit{jar} or \textit{classpath hell} (\url{https://wiki.c2.com/?ClasspathHell}), in reference to DLL hell.}, a common issue due to transitive dependencies, and (2) to enable automated upgrades---under-constraining allows the package manager to choose the ``latest stable'' version.

Automated upgrades (e.g. in Maven) usually take effect at build-time. But, the class loading mechanism used in runtime modularity frameworks like OSGi also supports underspecified version matching at runtime, allowing hot upgrades.

In the context of updating to remove vulnerabilities in libraries, tools that notify client developers about relevant changes have recently become more popular. These tools monitor vulnerability databases (primarily CVE) and notify downstream developers about available upgrades. This approach has been integrated into build tools for some ecosystems. Examples include \texttt{npm audit}\footnote{\url{https://docs.npmjs.com/cli/audit}}, \texttt{cargo audit}\footnote{\url{https://github.com/RustSec/cargo-audit}} and the audit plugin for Maven\footnote{\url{https://sonatype.github.io/ossindex-maven/maven-plugin/}}. These tools require proactive steps on the developers' part: they must explicitly include the relevant plugins and incorporate these plugins' feedback, making for a \textit{pull} model. A more aggressive \textit{push} model has emerged recently. Push-based tools integrate as services into GitHub and other hosting services and automatically propose upgrades to dependencies, often using the pull request mechanism. Examples of push tools include \textit{dependabot}\footnote{\url{https://dependabot.com/}}, \textit{renovatebot}\footnote{\url{https://github.com/renovatebot}} and \textit{snyk}\footnote{\url{https://snyk.io/}}. Under both the pull and push models, downstream developer workload increases proportionally with the number of (transitive) dependencies: even when a tool automatically proposes a security update, the developer remains responsible for ensuring that the change is actually safe, and weighing that against the severity of the reported vulnerability. \citeN{kikas17:_struc_evolut_packag_depen_networ} characterized dependency networks and how they grew over time in the JavaScript, Ruby, and Rust ecosystems, implying more update-related work for developers.

\paragraph{Example: automated upgrade woes.} A related problem to component upgrades arises in the context of server infrastructure running consumer operating systems such as Ubuntu. For consumer products, automated upgrades are critical for avoiding 0-day vulnerabilities; upgrading late leads to compromised client machines. Ubuntu therefore will auto-upgrade and reboot. But Ubuntu also markets itself to data centers for production use. However, professionally-managed server infrastructure is on a controlled upgrade cycle and does not need automatic upgrades or arbitrary reboots. In this context, uncontrolled upgrades provide all of the risk of breaking changes with few additional benefits. 

\subsection{Identifying and mitigating changes} In this section we have discussed upstream developers' responsibilities with respect to breaking changes, as well as the perils faced by downstream developers in upgrading late and early. Ideally, incompatible changes are flagged as such by upstream, typically with new major version numbers. Flagged changes require developers to understand whether the issues changes are applicable, while unflagged changes potentially require developers to debug problems caused by automatic upgrades. In both cases, downstream developers bear the ultimate responsibility for working around compatibility bugs that upstream has introduced.

One of our industrial colleagues reports that their company manages upgrades by hand because it is critical to manage breaking changes while incorporating needed updates. This reinforces the main thesis of this essay: developers (and other software users such as sysadmins) need better tool support to flag relevant changes in components so that they can painlessly apply changes at the appropriate time. In the remainder of this essay, then, we present a vision of the tools that ought to be created to help support developers manage upgrades.



	\section{Contracts: Documenting and Verifying API Behaviours}
\label{sec:about-contracts}

In the previous section, we discussed reasons that upstream makes
breaking changes, as well as the risks downstream developers take in
upgrading late (missing beneficial changes) and early (unnecessary
work).  Understanding the effects of upgrades reduces to understanding
the upstream components themselves. Usually this understanding is
abridged. Contracts are one way to express relevant properties of the
upstream components, for instance by specifying abstractions of
program behaviour. Components may change their contracts or their
conformance to these contracts.
Our
interest in contracts in this essay stems mainly from their relevance
as one tool for understanding software upgrades.

Even bugfix changes, where programs are modified to better conform to
their contracts, may be breaking changes in some cases (and we saw
real-world examples of this in Section~\ref{s:status_quo}).
Clients may rely on specific behaviours that do not meet the
specification (or, for that matter, any reasonable
specification). Emulation libraries such as the Windows emulator Wine
aim to preserve the original system's behaviour at a bug-for-bug
level\footnote{\url{https://wiki.winehq.org/Wine_Features}}.

Our thesis is that \emph{a critical part of helping developers safely upgrade is by
detecting changes to contract specifications and implementations}.
The question then arises: ``What exactly is a
  contract, and how do we find them?''  Of course, there are many ways to define
contracts! While
contracts are often thought of in terms of programming constructs
(e.g. types, pre/post-conditions, etc) they equally apply at higher
levels (e.g. REST APIs, protocols, etc) and lower levels
(e.g. linking) as well as to non-technical artifacts (e.g. licenses).
Contracts can cover both functional and non-functional requirements---a
significant degradation in the performance of a library
might be considered a contract violation.
Finally, contracts may be \emph{explicit}
(e.g. method {\tt foo()} returns a {\tt List}) or
\emph{implicit} (e.g. suggested by a method name such as {\tt export()}).

Contracts can help both upstream and downstream developers. From the
upstream perspective, the constraints imposed by the contract actually
represent freedom to change what is not in the contract. Behavioural
properties not mentioned in the contract may be changed. Of course, the provider of an upstream component must
preserve behaviour that is mentioned in the contract, or else
explicitly amend the contract. Non-functional properties
may be more fraught: there may be legitimate but undocumented expectations about 
performance, for instance. From the downstream perspective,
unchanged contracts, along with tools to guarantee conformance to
those contracts, can give developers more confidence that new versions
are safe to upgrade to; differences between contracts point out where
adaptation is needed.

In this section, we summarize related work in the areas of defining and verifying formal
and lightweight contracts, as well as work that detects contracts in code
that is not already annotated with suitable contracts (i.e. most
extant code).


\subsection{Formal Contracts}
\label{sec:formalcontracts}



In some communities, a function's contract is most commonly given
using the tools of preconditions and postconditions.  This evolved
from the early use of runtime assertions.  \citeN{Turing49a}~advocated
using assertions as stepping stones \emph{``from which the correctness
  of the whole program easily follows''}, and stylized assertions upon
entry to and exit from Pascal functions were proposed
by~\citeN{ILL75}.
Alphard~\cite{wulf76:_introd_const_verif_alphar_progr} was perhaps the
first programming language to provide explicit syntax for expressing
pre- and post-conditions as logical assertions~\cite{Rosen95}, a trend
continued by later languages such as
Turing~\cite{holt88:_turin_progr_languag} and
Eiffel~\cite{meyer92:_eiffel}.  The Java Modeling Language (JML)
provided a standard notation for specifying Java~\cite{LCCRC05}.

Of course, pre- and post-conditions are not the only means to specify
programs.  For example, using algebraic
specifications expressed as axioms~\cite{LZ75}, has received
considerable attention~\cite{ST97} and is arguably more
expressive~\cite{BH14}.  No matter---for our purposes, any means for
specify program behaviour is relevant (especially if machine readable,
such as CASL~\cite{ABBKMST02}) .

A common argument here is that safe substitution is possible when the
replacement artifact \emph{refines} the specification in question
(i.e. does not strengthen preconditions or weaken
postconditions)~\cite{JP01}.  Unfortunately, in practice, substitution
is only safe when the specifications are sufficiently complete. So,
the question is: \emph{how complete should functional specifications
  be?}  \citeN{MG14}~argue that \emph{``formal specifications need not
  encompass all requirements.  We can prove browser security without
  formalizing everything a web browser must do''}.  Likewise,
\citeN{BH95}'s second commandment of formal methods is \emph{``Thou
  shalt formalize but not over formalize''} which arises from the
inherent cost of applying formal methods.  \citeN{PFP13}~capture the
issue succinctly:

\begin{quote}
  \emph{``But what about strong specifications, which attempt to
    capture the entire (functional) behavior of the software? Should
    we dismiss them on the grounds that the effort required to write
    them is not justified against the benefits they bring in the
    majority of mundane software projects?''}
  \end{quote}


Our position, in line with the above, is that even if (functional) specifications were available,
one could not expect them to be complete. A key open challenge, then, is ensuring that
relevant contracts are sufficiently complete for the purpose of detecting breaking changes and yet still usable. Usability includes being concise enough and being amenable for reasoning about changes therein, among other considerations.
The general challenge of ensuring that software conforms to its contracts also remains important.

\paragraph{Non-functional properties.}
We reiterate that performance degradations can constitute
breaking changes between versions and yet are normally considered ``non-functional'' properties. The notations for formal
contracts highlighted above---Turing, Eiffel, and JML---focus
on functional specifications, and thus omit relevant non-functional
specifications (e.g. Worst-Case Execution Time).
Instead, upstream components often use free-form text to document performance
expectations: Java's {\tt
  List} API documentation requires that some operations must
take linear time, and its subclass {\tt ArrayList} describes more
specific performance expectations for some methods.


\subsection{Lightweight Contracts}

While contracts written in first-order logic offer considerable
expressive power, they are costly to write, check and maintain.
Typically, they are only used in extreme situations, such as 
safety-critical systems~\cite{CS14,DELMM14}.
We consider more lightweight approaches to also be a form of code contract.
These approaches can live at a more accessible cost/benefit tradeoff.
Indeed, \citeN{RHDB08} observe
that \emph{``type systems are the most successful and widely used formal
  methods for detecting programming errors''}.
We believe that more lightweight contracts are also an important tool in helping to
detect breaking changes: even if they are less expressive
than heavier-weight contracts, there are more of them available to use.

Numerous systems enrich the expressiveness of existing type systems
with contract-related concepts.  Examples include systems for
\emph{non-null types}~\cite{FL03}, \emph{ownership} and
\emph{uniqueness types}~\cite{CPN98}, and others~\cite{FTA02,VPEJ15}.
Indeed, the designers of \textit{Spec\#} chose to mix lightweight
static analysis for checking non-null types and method purity with
automated theorem proving for functional specifications given in
first-order logic~\cite{BFLMSV11}.  Likewise, modern languages are
beginning to encode such contracts in their type systems
(e.g. ownership in Rust)~\cite{RustBook}.  

The \emph{Checker Framework}~\cite{papi2008practical} is a popular pluggable type system  
for Java which comes with several checkers (nullness, taint, regex, etc.). It also allows third parties
to write additional checkers. 
The \emph{Checker Framework} uses standard Java type annotations. For instance, \texttt{@NonNull String} is considered a subtype of \texttt{String} which is guaranteed to store non-null references only. Checks are performed at compile time. To benefit from the \emph{Checker Framework}, developers must expend some effort to annotate the program. However, Maven Repository usage statistics for the artifact defining standard \emph{Checker Framework} annotations\footnote{\url{https://mvnrepository.com/artifact/org.checkerframework/checker-qual}} suggests that developers responsible for hundreds of projects have chosen to include Checker annotations.

Method \emph{purity} is an interesting case here for several reasons:
firstly, checking purity is well within the capability of lightweight
static analysis (as long as annotations are
provided)~\cite{SR05,HM12,NNRM15}; secondly, changing the purity
status of a method could certainly be considered a breaking
change---and yet is rarely considered in practice.  One can easily
imagine tools such as \textit{Revapi} being extended to spot changes
in purity using an {\tt @Pure} annotation and an accompanying
intraprocedural checker (e.g. \textit{JPure}~\cite{Pea11}).  {\em
  Computational contracts} (like effects~\cite{GL86}) take this further by
allowing one to specify, for example, whether a given method {\em must
  not} invoke some other method~\cite{STD15}.


The simplest form of contract is the set of exported type
signatures for public methods (along with the informal expectations on behaviour implied by
method names). Tools like \textit{clirr} and \textit{Revapi}, mentioned
earlier, evaluate version compatibility using type signatures, and the
elm programming language's ecosystem tools ({\tt elm bump})
automatically increase version numbers when they observe incompatible
changes in a package's exported
types\footnote{\url{https://github.com/elm-lang/elm-package/blob/master/src/Bump.hs}}.
\textit{rust-semverver} performs a similar check for Rust crates.
We believe that such tools would benefit from a deeper understanding of
the code---one that could be obtained from contracts like those
described above.

\subsection{Detecting Contracts}



An unfortunate challenge with the status quo is that, for the vast
majority of existing code, most contracts are implicit (or only
described in natural-language documentation).  \citeN{AM02}~formulated the \emph{Closet Contract Conjecture} as a means to explain
this:

\begin{quote}
\emph{``Because the benefits of contracts are so clear to those who use them, it's natural to
suspect that non-Eiffel programmers omit contracts because they have no good way to
express them, or haven't even been taught the concepts, but that conceptually
contracts are there all the same: that inside every contract-less specification there is a
contract wildly signaling to be let out.''}
\end{quote}

The basic assumption here is that, despite language limitations,
programmers will encode some contract information \emph{using whatever means they have
  available}.  For example, programmers might use exceptions to check pre- and
post-conditions. Or, they might exploit mechanisms for maintaining invariants
over state (e.g.\ encapsulation).  Indeed, a large study of existing
code found some use of contracts beyond types~\cite{DPZB17}.

However, the argument of \citeN{PFP13} that \emph{``programmers are
  willing to write specifications if it brings tangible benefits to
  their usual development activities''} makes one question whether
programmers will ever be sufficiently motivated to provide
sufficiently detailed code contracts.  \citeN{L00} made the
following observation:

\begin{quote}
 \emph{``Although annotations capture programmer design decisions and
   provide a stylized way to record these, the reluctance to cope with
   the burden of annotating programs remains the major obstacle in the
   adoption of extended static checking technology into
   practice.''}
 \end{quote}

Likewise, reflections on the \textit{Spec\#} project included observations
around the difficulty of converting existing code to include
contracts~\cite{BFLMSV11}.  Non-Null by Default~\cite{CJ07} suggests
one possible adoption strategy, but may be difficult to generalize to
other properties.

We therefore consider the question of how one can identify
contracts hidden within legacy code.
 
\paragraph{Formal contract inference.}


The traditional problem of \emph{specification inference} ---that is,
the automatic inference of \emph{pre-} and \emph{post-conditions}---from
legacy code has been studied extensively~\cite{PCM09,CCL11}.  Indeed,
\citeN{L00} argues this is crucial to enabling more wide-spread uptake
of verification technology:

\begin{quote}
\emph{``For programming teams with large amounts of already written
  code, the initial investment of adding annotations to the legacy
  code seems daunting.''}
\end{quote}


Most approaches to this problem are based on some combination of
inferring \emph{weakest preconditions} and \emph{strongest
  postconditions}~\cite{Dijkstra76,BL05}. \textit{SnuggleBug} is a
salient example which employs a range of techniques~\cite{CFS09}.  For
example, it interleaves symbolic analysis with call graph construction
to yield a more precise construction, and employs generalization to
ensure specifications contain only ``pertinent'' information.


Program documentation (using natural language processing techniques) is also a source of program properties; for instance, \citeN{yang18:_towar_extrac_web_api_specif_docum} extract endpoints from documentation of web APIs. Such approaches also work for extracting models from documentation~\cite{zhai16:_autom_model_gener_docum_java_api_funct} or from code comments.

Specification inference can also extract properties from program executions as in Daikon~\cite{Ernst2000:PhD}. The work of \citeN{DBLP:conf/sigsoft/McCamantE03} was particularly foresighted and relevant here: they developed an analysis based on dynamic invariant detection for predicting component upgrade problems in the context of a particular client. This essay proposes a research programme which is broader than that work on a number of axes: among others, we include the unknown-client case; propose the use of static and hybrid analyses; and point out the importance of non-behavioural properties.

\paragraph{Lightweight contract inference.}


While inferring full specifications might be considered an ideal (in
some sense), such specifications are expensive and fail to realize
their full benefits without tools which can statically check them.  As a result,
much work has focused on inferring properties that can be viewed
either as limited contracts or extended type systems.  For example,
researchers have developed tools that infer non-null
annotations~\cite{KH07} (in a world without Non-Null by Default) and
ownership annotations~\cite{FF07,ML09,HDME12,Pea13d}.

We believe that all of these inferred specifications (whether heavyweight
pre/postconditions or lightweight type annotations) can help in detecting
breaking changes. After all, any change in an inferred specification implies
some potentially-relevant change in the implementation.


\subsection{Checking Contracts}

Inferred contracts clearly reflect their implementations (but can
be in danger of over-fitting).  And, for declared contracts to
be useful in detecting breaking changes, implementations must conform
to their declared contracts (which must also be sufficiently
complete).  Much effort has been devoted to developing tools which
statically check implementations against contracts.  Indeed, after the
success of early prototypes~\cite{LGHKMOPS95,DLNS98,FLLNSS02} work
continues apace to develop mature and practically useful
tools~\cite{BFLMSV11,Leino10,Leino12,FP13,JSPVPP11,Pea15c}.
Pragmatically, however, such tools remain unreliable and difficult to
use~\cite{BFLMSV11}.  For example, \citeN{GHHJX14} found that, despite
their considerable resources, it was \emph{``not feasible to produce a
  rigorous formal proof of correctness [\ldots of] a file system''}.
As such, we must accept that the vision of software development
routinely using verification tools remains largely unfulfilled and
that, perhaps, the initial enthusiasm surrounding Hoare's verification
Grand Challenge~\cite{Hoare03} has been replaced by a more gritty
reality.

In the language of contracts, a change $A \rightarrow A'$ is breaking, that is, version $A'$ is incompatible with $A$, if: (1) the (relevant) contract for $A'$ does not imply the contract for $A$; or (2) the implementation of $A'$ does not fulfill the (relevant) contract for $A$. Contract changes can still be breaking even in the absence of implementation changes; consider, for instance, a license change that revokes permissions to use a component.

Despite these observations, we argue that formal contracts---\emph{even without static checking}---still offer significant potential in the
battle to detect breaking changes.  For example, they could help
dynamic analysis tools find witnesses to breaking changes by providing
an oracle to work from.  \citeN{Whit00} notes \emph{``Without a
  specification, testers are likely to find only the most obvious
  bugs''}---something which applies equally well to breaking
changes.



We also argue that important differences in perspective exist when
contract checking as an upstream developer versus as a downstream
developer.  This is because upstream developers care about whether
changes can adversely affect \emph{any} of their clients (e.g ensuring
minor changes are not breaking).  In contrast, downstream developers
care only about whether \emph{they} are affected by upstream changes.

For instance, \citeN{mostafa2017experience} employ 
\emph{cross-version testing} to find breaking changes in Java libraries.
For a given version pair, this checks whether the existing regression
tests still pass in the updated version.  In a sense, these regression
tests serve as placeholders, representing existing downstream clients,
and can reveal non-trivial incompatibility errors.  Indeed, some of
the examples discussed in Section~\ref{sec:upstream} were detected
using cross-version testing.  
\textit{Differential regression testing} is a similar approach, 
applied to services~\cite{godefroid2020differential}.

Focussing on a single client at a time, the work
of~\citeN{mezzetti2018type}, which introduces \emph{type regression
  testing}, helps upstream developers avoid creating breaking changes
in their node.js libraries.  They leverage the known dependencies of
the library under analysis and use the dependencies' test cases to
build a model of that library---in particular, of how the library is
actually used downstream.  Then, comparing models before and after a
change detects changes that are breaking with respect to one client.
Looking across clients,
\citeN{mujahid20:_using_other_tests_ident_break_updat} pool tests from
a range of clients to determine whether a particular library change is
generally breaking.

In what follows, we consider a range of pragmatic approaches which
could be used by both upstream and downstream developers to detect
breaking changes.  While applicable in both scenarios, it is useful
to remember the differences in perspective here.  For example, a
breaking change in an \emph{upstream} position may not be breaking in
the \emph{downstream} position (e.g. because the client doesn't use any
of the affected method(s) in question).




\paragraph{Dynamic checking.}


If static checking of certain contracts remains beyond the state-of-the-art,
then dynamic checking is the pragmatic choice.  Indeed, tools for
specification-based testing have more than proved their worth in
recent times~\cite{SC96,BBP96,SR11,ABL14}.  The introduction in .NET
4.0 of \emph{Code Contracts} checked at
runtime~\cite{BFLMSV11,CodeContracts} is perhaps illustrative of such
an approach.  They helped push contracts into the mainstream
while, at the same time, giving breathing space for work to
continue on enabling the static
verification of contracts~\cite{faehndrich10:_static_verif_code_contr}.

\textit{QuickCheck}~\cite{CH00b} provides a concrete example which highlights
the potential here.  This tool automatically generates tests based on
user-provided specifications and, while originally developed for
Haskell, has subsequently been implemented for other languages
including Erlang, Java and C~\cite{AHJW06,CPSHSAW09}.  More
importantly, \textit{QuickCheck} has demonstrated value on industrial-scale
projects.  For example, it found hundreds of problems in over 1MLOC of
low-level C for Volvo~\cite{Hughes16}).
The Java Modeling Language (JML) provides another case in point.
While mature tools for static checking JML contracts have remained
stubbornly elusive~\cite{FLLNSS02}, others have found success by
stepping back from this ideal and employing dynamic techniques
instead.  For example, several tools exploit JML contracts for
automated testing, including \textit{JMLUnit}~\cite{CL02},
\textit{JMLUnit-NG}~\cite{ZN10}, \textit{Korat}~\cite{BKM02}, \textit{JMLAutoTest}~\cite{XY03},
\textit{TestEra}~\cite{DM04} and more~\cite{BDL06,CR08}.

\emph{Randoop} is another illustrative example~\cite{PLEB07,PE07}.
It applies random test generation to Java but, in this case, the
lack of an obvious test oracle presents a significant obstacle.  For
example, while throwing a {\tt NullPointerException} could be
considered a test failure, it could equally be considered acceptable
behaviour in some circumstances (e.g. a method's preconditions were not met).  Like \textit{QuickCheck}, \textit{Randoop} relies on
user-defined contracts to clarify what should and should not be
considered correct behaviour.  As a shortcut, it provides various built-in
contracts (e.g. computing a {\tt hashCode()} must not
throw an exception).

Finally, some tools do indeed operate successfully in contexts with
limited contract information without requiring user
intervention~\cite{WK18,LS18}.  An excellent example would be 
\emph{American Fuzzy Lop (AFL)}, which has proved effective at generating inputs
which crash programs (e.g. by causing segmentation faults).  The
tool employs an evolutionary algorithm which attempts to increase
coverage by mutating test inputs.  From the perspective of this paper,
however, it is questionable as to whether \textit{AFL} constitutes a tool for
dynamic contract checking. Unlike the other tools highlighted
above, \textit{AFL} gives no indication as to what contract was violated or
where this occurred other than checking for the implicit contract that no crash should occur.  
In some sense, one might instead classify \textit{AFL} as
\emph{contract-oblivious}. As such, it would seem to offer limited
utility as part of a semantic versioning calculator. Concolic
testing, e.g.~\cite{SMA05}, shares the disadvantages of
test generation tools in the context of contract verification.

\subsection{Recent Trends in Static Analysis}
Lightweight contracts lie at the boundary between the type systems and
static analysis communities. On the type systems side, the \textit{Checker
Framework} mentioned earlier enables developers to add and verify
custom type annotations in their code, potentially using specialized
checkers.  Moving towards static analysis, ownership annotations are
viewed as types but verified with techniques closer to static analysis
(as demonstrated by Rust's flow-sensitive borrow checker).  Finally,
at the other extreme, interprocedural static analysis can identify that
methods are \emph{side-effect free} or \emph{pure}~\cite{CBC93,SR05}.
Such analyses are useful to reason about the evolution of depended-on components: 
changes to those properties as components evolve can introduce behavioural changes 
that can break clients.

Any static analysis integrated into modern build processes
must be sufficiently fast and accurate. But it is challenging to build analyses that are both fast and 
accurate.
First, the performance of
sophisticated interprocedural static analyses, like pointer analysis
and callgraph construction, is often poor. This is a consequence of
the high complexity of the algorithms being used and the large problem
sizes. \citeN{reps1998program} dubbed this the cubic
bottleneck. Second, static analysis lacks important information
available only at runtime, and making conservative assumptions
introduces false negatives, i.e. reduces recall~\cite{livshits2015defense,suirecall}.

Two recent trends addressing these limitations hold great promise and may apply especially to reasoning about change.

\paragraph{Incremental and scalable static analysis.}
As software lifecycle iterations become ever shorter and programs ever larger, novel static analysis techniques have been deployed to make timely analysis possible. The elegant solution is to incrementalize---moving from computing results from scratch on each analysis run to updating the static analysis models and reading results from the models. New techniques to incrementalize are emerging for static analysis frameworks like \textit{Doop}~\cite{bravenboer2009strictly}, \textit{bddbddb}~\cite{whaley2005using}, and \textit{Flix}~\cite{madsen2016datalog} which are based on Datalog-based representations. The commercial Semmle tools also employ Datalog (QL)\footnote{\url{https://help.semmle.com/QL/}}. In Datalog, adding information and reading off new incremental results is trivial, as the fixpoint can be easily recomputed. The situation is significantly more complex when information has to be retracted. There is recent research in this space, with the incremental \textit{DDlog} engine recently becoming available~\cite{ryzhykdifferential}. The \textit{Doop} repository contains an experimental analysis based on \textit{DDlog}\footnote{\url{https://bitbucket.org/yanniss/doop/}}. There is active work to add incremental computation to \textit{Souffl\'e}, the default engine used by \textit{Doop}~\cite{zhao2019incremental}.

As another, particularly relevant, example,
the Facebook \textit{Infer} tool~\cite{ohearn17:_findin_infer}
performs inference
and inter-procedural verification of program properties that can be
viewed as extended types, including nullness and resource leaks. It
uses a compositional shape
analysis~\cite{calcagno11:_compos_shape_analy_means_bi_abduc}.
\textit{Infer} is extremely scalable---it runs quickly enough even
on codebases with up to tens of millions of lines---even though it performs
an inter-procedural analysis. One key to its scalability is its use of incremental
static analysis: it records analysis information on each run and reuses this
information on subsequent runs.
\textit{Infer} shows that (with sufficient engineering effort) such tools have
the potential to be usable on industrial-sized codebases.

\paragraph{High-recall static analysis.}
Traditionally static analysis has focused on balancing precision and scalability. More recently, researchers have turned their focus to recall or soundness, i.e. addressing issues around false negatives~\cite{livshits2015defense}. \citeN{suirecall} have demonstrated the  importance of this---standard  call graph construction algorithms suffer from significant numbers of false negatives, undermining the utility of  graph- and points-to-based analyses.  Moreover, there are multiple sources of unsoundness, each having a significant impact, while research has focused on only one category (reflection)~\cite{livshits05:_reflec_analy_java,smaragdakis2015more}.

Our interest in this essay is in detecting breaking changes. Recall is relevant to this application because a sound, or at least a high-recall, analysis means that there are either no issues being missed, or that the number of issues that are being missed is reduced to an acceptable level. This can give users enough confidence to use tools to detect compatibility-related bugs. High-recall tools avoid or minimize chances for expensive runtime errors caused by undetected bugs, and the corresponding need to roll back upgrades. 

Recent progress in this field has focused on two areas: modelling of various dynamic language features in pure static analyses~\cite{fourtounis2018static,fourtounis2019deep}, and hybrid analyses blending information gathered from program runs (via stack sampling, instrumentation, or heap snapshots) into static analyses~\cite{bodden2011taming,grech2017heaps}.







	\section{API Surfaces: Which Upgrades Are Relevant?}
\label{sec:api-surfaces}

An upstream developer makes available some API for their component and is responsible for maintaining it. However, almost always, when a client uses a component, it does not interact with the entire component.  If interaction is cast in terms of method invocations and field accesses, then the interaction is usually restricted to a subset of the component's methods and fields---the component's API surface. A particular downstream developer is only affected by upstream changes in the part of the API surface that they use. The question thus arises:
what constitutes a component's API surface?



\subsection{The Open World Assumption} The easiest approach is to include all of the component's public fields and methods in the API surface. Furthermore, clients can also access protected methods and fields through subtyping, so they should be included as well. And, since reflection can be used to bypass encapsulation boundaries, all methods and fields are accessible in principle. We can refer to this maximalist approach (with any of the variations discussed) as the \textit{open world assumption}: everything must be considered to be part of the API surface because some client out there in the open world might use and depend on it.

In many cases, the open world assumption would lead to precision issues: analyses based on such an open world assumption would lead to incompatibilities being detected in parts of components that do not have an impact on any client. The extreme maximalist position which includes reflective accesses, in particular, would declare practically any nontrivial change to a component to be incompatible. This is reflected in the statement by Martin Buchholz quoted in the introduction: ``Every change is an incompatible change.'' But can we do better?


\subsection{Limiting the API surface} One idea is to retain the open world assumption, but to add a declaration which restricts the published API surface. This is usually done either by declaring that the component exports certain artifacts, or by declaring that the component publishes certain interfaces and implements services that implement these interfaces. Early successful examples are CORBA IDL modules, OSGi bundles and its various service extensions, and the service loader mechanism in Java widely used in JDBC 4 drivers. A more recent attempt to standardize this as part of the Java language and runtime is Java 9 Modules (JSR 376, formerly known as project jigsaw)~\cite{jsr261}. Java modules can declare both a list of exported packages and services (which can be outside the exported packages), and can also restrict reflective access to module internals. While these restrictions are enforced at both compile-time and at runtime, clients can still circumvent encapsulation boundaries\footnote{The \textit{--add-exports} option described in ~\cite{jsr261} allows the client to declare that a depended-upon component's public API is wider than published by the component authors.}. Circumvention measures are intended to be used temporarily to assist projects with the transition to modules rather than permanently.


\begin{table}[]
\small	
	\begin{tabular}{l|r|r|r|r}
		project                   & \multicolumn{2}{c|}{all} & \multicolumn{2}{c}{module} \\ 
		& methods     & fields     & methods       & fields      \\ \hline
		log4j-2.12.1              		& 2,259        & 452        & 2,252          & 449         \\ 
		org.glassfish/jsonp-1.1.2 & 575         & 128        & 36            & 1           \\ 
	\end{tabular}
	
	\caption{\label{tab:apisurface}API surface size of Java libraries can vary widely: all non-synthetic public and protected fields and methods, vs exported public and protected methods and fields.}
\end{table}

\paragraph{Preliminary experiment: API surface width} We implemented a simple bytecode analysis to further investigate API surfaces. Table~\ref{tab:apisurface} shows the extent of potential versus declared API surfaces in two example libraries. The libraries analyzed are both widely used---log4j is a popular logging framework, while  glassfish jsonp is an implementation of the jsonp JSON parser API. Both provide explicit Java 9 module definitions in the form of a \texttt{module-info.java} file. Our results indicate that adding a module definition makes little difference to the size of the log4j  API surface---classes in this library implementing the various appenders, configuration, levels, etc. are designed to be accessible by clients to configure logging.  We can expect a similar situation for many other libraries that are collections of useful utilities, like guava or Apache commons collections: these components have a deliberately wide API surface and relatively less code behind the surface. The results for {org.glassfish/jsonp-1.1.2} paint a very different picture: module information leads to a dramatic reduction of the API surface size. The {jsonp} component exports only one package (namely, \texttt{org.glassfish.json.api}) and one implementation class (\texttt{org.glassfish.json.Json\-ProviderImpl}). {jsonp} provides a well-defined single piece of functionality, backed up by a complex, but well-encapsulated, implementation.
Of course, clients using this component will still need to execute (at least some of) the encapsulated code, and are therefore sensitive to changes. We refer to this as \textit{indirect API access}.

\subsection{Indirect API Access} 

The discussion so far has focused on APIs directly accessible by clients, i.e. the types, methods and fields that can be referenced in client code. But this ignores the parts of components exposed to clients due to the data flow resulting from pointers and dynamic dispatch. Equivalently, in functional programming terms, clients can get access to program-internal parts of the program through lambdas returned by invoked components. A similar situation arises in languages with function pointers when such pointers are returned to clients, or in callback-oriented programming popular in languages like  JavaScript. 

Consider for instance {yasson}, a Java framework which provides a standard binding layer between Java classes and JSON documents\footnote{\url{https://github.com/eclipse-ee4j/yasson}}.  {Yasson-1.0.1} depends on {org.glassfish/json-1.1.2}, the reference implementation and default provider for JSR 374 (JSONP). None of the classes in {yasson} reference any types belonging to a \texttt{org.glassfish.*} package. Instead of using such types, {yasson} uses the factory method \texttt{javax.json.spi.JsonProvider.provider()}, which uses a combination of service loader lookup based on provider library metadata plus a default reference (which happens to be \texttt{org.glassfish.json.JsonProviderImpl}, if that is in the class path). However, when executing {yasson}\footnote{We used instrumentation to track method, constructor and field use, using a Maven build (\texttt{mvn test}) as driver.}, it turns out that 87 methods are invoked by the client (test driver), of which only 2 are declared in a class that is part of the exported module API. Field accesses (71 fields read and 84 fields written) and allocations (28 constructors invoked) also happen almost exclusively in non-exported classes and packages, through methods on returned objects. This is hardly surprising, and widely accepted programming techniques mandate exactly this: simple APIs (\texttt{JsonProvider}) hiding complex application internals ({glassfish}). But as soon as we try to bring the semantics into interface contracts, this starts to matter: if a new version of some private, indirectly invoked method suddenly fails with a \texttt{NullPointerException}, that exception is propagated across component boundaries into clients, and constitutes a breaking change.

This illustrates the need for ``deep'' program analysis (including pointer analysis) to check interface implementations. Restricting the API with modules  or similar is still useful as it can be used to restrict the scope of the analysis by providing meaningful analysis drivers, making it faster, and more accurate. Similar analyses have been proposed for vulnerability detection in service-based systems including Java modules~\cite{dann2019modguard} and OSGi~\cite{goichon2013static}.

\subsection{Investigating a Closed World} 

A more radical idea is to abandon the open world assumption completely and adapt a\textit{ closed world assumption} based on a usage analysis with actual clients. For instance, \citeN{mora18:_clien} have developed the \textit{Clever} tool which verifies equivalence of two library versions given a specific client. While it is generally not possible to know all clients\footnote{An exception are Java modules which have the unusual feature of allowing components to restrict the clients that they export to.}, for popular components, many clients are known through tracking dependencies in repositories.  Restricting the API surface with module definitions can be seen as a step to specify \textit{intended uses}. Limiting the set of anticipated clients, on the other hand, is a step towards reasoning about \textit{actual uses}, or to be precise, an approximation of use by actual clients.  While usage analyses are tempting, implementing them is challenging. For instance, a reasonable approach would be to use static call graph analysis to detect which library methods are invoked by clients, similar to the approach used in~\cite{hejderup2018software}. This has the usual static analysis issues related to recall and precision. A dynamic analysis is also possible, for instance by running client tests, instrumenting components, and recording invoked methods and accessed fields, similar to the approach used in \cite{mezzetti2018type}. Unfortunately, dynamic approaches are unsound and may produce an under-approximation of the actual API surface. 

\subsection{Integrating API Concepts into Programming Language Design}

As discussed earlier in this section, understanding module interfaces requires surprisingly sophisticated program analyses. Programming language design can have a key role to play in enforcing modularity and hence in preventing breaking changes. Many of today's mainstream languages were originally designed in the 90s, when systems were smaller and there was less reuse---certainly not the repositories easily available today. Accordingly, encapsulation boundaries focused on sub-component groupings like packages and namespaces (in Java and C++/C\# respectively), which resulted in many classes and methods being declared as public, in turn inflating publicly accessible APIs.

Today, however, as reuse (in the form of vast component ecosystems) and rapid automated evolution continue to gain popularity, language designers can help developers by integrating API-related concepts into languages as first class citizens, as seen for instance in Java 9 modules. It is particularly hard to get the level of granularity right: what constitutes a component, what are the component boundaries, and what is exposed. Java prior to version 9 demonstrates this: without an explicit module system, encapsulation boundaries were defined based on packages and classes, and to a lesser extent, on the class hierarchy. This resulted in the boundaries being required to include large numbers of public classes and methods. Many of those public classes and methods were supposed to be component-internal, but their encapsulation could not be enforced with language constructs. .NET also previously had public APIs defined in namespaces labelled \texttt{Internal}, which were only encapsulated by convention and not by the compiler.
Integrating appropriate features into languages to define component APIs has the immediate advantage of making components easier to test and (statically) analyze, and therefore also easier to maintain. Analyzability and testability are core components of the maintainability category in ISO/IEC 25010:201~\cite{iso2011iso}. In particular, reasoning about change becomes easier as the required scope of the analyses becomes smaller. Language support can thus help upstream developers make fewer accidental breaking changes and downstream developers import fewer such changes.

        \section{Towards A Taxonomy of Breaking Changes}
\label{sec:taxonomy}

Empirical work on evolution-related practices in different communities~\cite{bogart2016break,decan2019package,dietrich2019dependency} suggests that cultural norms greatly differ across communities when it comes to assigning version numbers, handling API stability, and evolving and adopting semantic versioning. For instance, \citeN{xavier17:_histor_impac_analy_api_break_chang} analyzed breaking changes in Java libraries hosted on GitHub using a tool to find changes that they identified as breaking; they found that about 15\% of changes were breaking but that a median of 2.5\% of clients were impacted by such changes. Tool support is crucial to enable faster, less error-prone dependency upgrading. But if tools are to have impact beyond academia and in practice, they need to respect distinctive community practices and facilitate community workflows.

Linters, starting with the classical \textit{lint} tool~\cite{johnson1977lint}, along with lightweight static analysis tools like \textit{FindBugs}~\cite{hovemeyer2004finding} and its spiritual successor \textit{SpotBugs}, show how tools can succeed.
\textit{FindBugs} employs relatively simple (usually intra-procedural) static checks. Its checks are categorized and associated with severity and confidence values. This information helps users select the most relevant checks and interpret the results.
In part because of this categorization, \textit{FindBugs} has been successfully deployed in both open source and industrial projects~\cite{ayewah2008using}.

At its core, the categorization of checks in \textit{FindBugs} is based on a taxonomy of issues. Taxonomies for evolution-related issues exist, going back to the seminal work of \citeN{beugnard1999making}. While \citeauthor{beugnard1999making} provided a high-level classification that can be a  useful starting point, more details are needed. Specifically, relevant to this essay's thesis, refined concepts are needed to describe API evolution and compatibility in statically and dynamically typed languages.

\begin{figure}[!t]
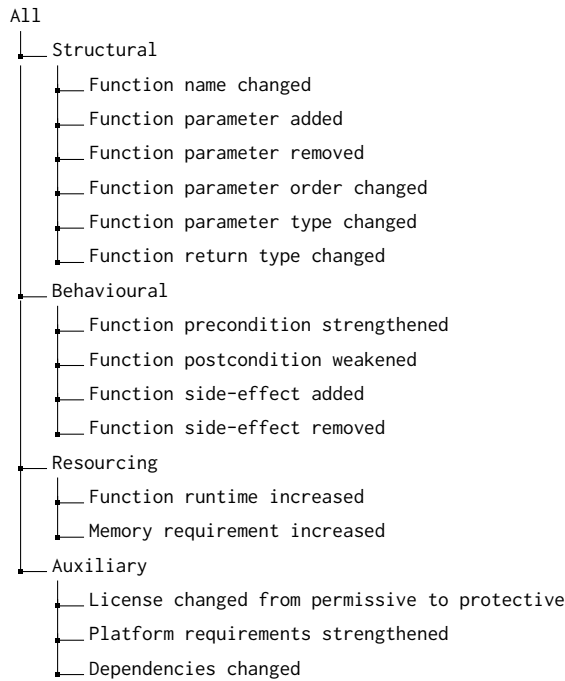

\begin{minipage}{\columnwidth}
	\begin{footnotesize}
\dirtree{%
	.1 All.
	.2 Structural.
	.3 Function name changed.
	.3 Function parameter added.
	.3 Function parameter removed.
	.3 Function parameter order changed.
	.3 Function parameter type changed. 
	.3 Function return type changed. 
	.2 Behavioural.
	.3 Function precondition strengthened.
        .3 Function postcondition weakened.
        .3 Function side-effect added.
        .3 Function side-effect removed.
	.2 Resourcing.
	.3 Function runtime increased.
	.3 Memory requirement increased. 
	.2 Auxiliary. 
	.3 License changed from permissive to protective. 
	.3 Platform requirements strengthened. 
	.3 Dependencies changed.
}
        \end{footnotesize}
\vspace*{1em}        
\end{minipage}
\caption{An example (incomplete) taxonomy for breaking changes.}
\label{f:taxonomy}
\end{figure}

Figure~\ref{f:taxonomy} illustrates concepts that might belong to a high-level generic taxonomy, primarily at a per-function level. It is not intended to be complete.

Such a generic taxonomy can serve as a guide for defining language-specific taxonomies which take language characteristics into account. Some languages already have catalogues of incompatible API changes, notably Java~\cite{rivieres20evolving}, .NET~\cite{msft:_break_chang,msft:_break_chang_rules}, Haskell\footnote{\url{https://pvp.haskell.org/}} and Rust\footnote{\url{https://github.com/rust-lang/rfcs/blob/master/text/1105-api-evolution.md}}. These catalogues fit in with our the generic notions of compatible and incompatible change patterns. For instance, in the context of static typing and subtyping, our notion of changes to function (method) return types includes 1) specializing return types (which are often compatible changes), and 2) any other changes (incompatible). For both Java and .NET, changes can further be classified by the time when incompatibility problems arise: compile-, load- or run-time.  Specializing the return type of a function is a source-compatible change, but will break binaries during loading and linking as it is not binary compatible.  A taxonomy will need to take such modalities into account.

Categorizing certain changes to fit into top-level categories is not always straightforward---changes can span categories. For instance, when the resource requirements of a component increase, then so do the resource requirements of the downstream application. But this increase could also trigger downstream behavioural changes, e.g. timeouts and out of memory errors. Classifications are still useful in understanding changes, but are not a panacea.

At the community level, a taxonomy can help 1) inspire members to build tools to reason about change and 2) form policies to concisely express social contracts within those communities.


	\section{Towards Declarative Semantic Versioning}
\label{sec:version-numbers}
If we consider the upgrade process as a collaborative activity that involves both upstream and downstream developers, then the semantic versioning specification from~\cite{preston2013semantic} assigns the cost of upgrading almost entirely to the upstream developer, who must decide on a version number for a new release. If that is properly done, then the downstream developer can easily automate the upgrade process, perhaps with some additional checks to catch remaining issues resulting from the shortcomings of the (upstream) version calculation.

	\tikzstyle{block} = [rectangle, draw, 
	text width=5em, text centered, rounded corners, minimum height=2em]
	\tikzstyle{bt} = [rectangle, draw, 
	text width=1em, text centered, rounded corners, minimum height=2em]
	\usetikzlibrary{calc}
	\usetikzlibrary{arrows.meta}
	
	\begin{figure*}[h!]
		\begin{tikzpicture}
		\node[bt] (A) {$A$};
		\node[bt,below of=A] (A') {$A'$};
		\node[block, right of=A, text width=3em, xshift=1.5em] (Tool1) { Tool 1 };
		\node[block, right of=A', text width=3em, xshift=1.5em] (Tool2) { Tool 2 };
		\node[block, right of=Tool1, text width=5em, xshift=3em] (F1) { Structural Facts };
		\node[block, right of=Tool2, text width=5em, xshift=3em] (F2) { Behavioural Facts };
		\node[block, above right of=F2, text width=5em, xshift=5em] (R) {Reasoning};
		\node[block, right of=R, text width=5em, xshift=4em] (I) {Impacts $A \rightarrow A'$ {\bf Breaking}};
		
		\draw[->] (A) -- (A') {};
		\draw[-Latex] let
		\p1 = ($ (A') - (A) $),
		\n2 = {veclen(\x1, \y1)*.5}
		in
		(A)++(0,-\n2) -- (Tool1);
		\draw[-Latex] let
		\p1 = ($ (A') - (A) $),
		\n2 = {veclen(\x1, \y1)*.5}
		in
		[-Latex] (A)++(0,-\n2) -- (Tool2);
		\draw[-Latex]  (Tool1) -- (F1);
		\draw[-Latex]  (Tool2) -- (F2);
		\draw[-Latex]  (F1) -- (R);
		\draw[-Latex]  (F2) -- (R);
		\draw[-Latex]  (R) -- (I);
		\end{tikzpicture}
		\caption{Parts of a semantic version calculator. Analysis tools analyze the differences between versions $A$ and $A'$ of a component, producing
			facts which can be manipulated by reasoning tools. Developers can read off the impacts of their changes,
			including whether the change is breaking or not.\label{fig:calculator}}
		
	\end{figure*}
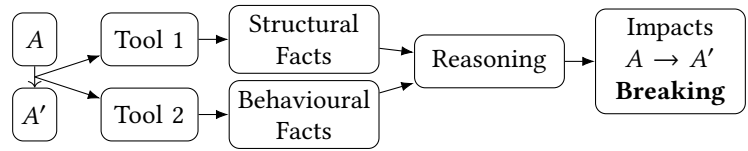

However, relying on the best intentions of upstream developers is problematic. Better tools might help. However, even improved program analysis tools remain subject to both false positives and false negatives, inevitably resulting in version numbers that are sometimes too optimistic or too pessimistic. We advocate for tools that use taxonomies as a way to significantly improve the situation. Using contracts, such tools could calculate and communicate impacts of changes in a way that is palatable to developers.

We view a taxonomy of changes as a foundation for principled semantic versioning. Both upstream and downstream developers can (semi)formally define their own semantics for a breaking change by including some taxonomy categories and not others. In all cases, tools can perform automated checks and report results using the categories of the taxonomy. An upstream developer can use output from a version calculator tool in two ways: 1) to guide the selection of changes to be included in a release; and, 2) to signal the impacts of their changes (usually through the semantics of version numbers, but also through change logs). A downstream developer can verify proposed upgrades to their dependencies, verifying that the published changes are indeed acceptable. We have discussed how different developers and developer communities, as downstream consumers, are differently motivated by backwards compatibility and weight the severity of potential changes differently.

Although we classify changes as either compatible or incompatible, not all incompatible changes are equally severe. Taxonomies can help both upstream and downstream developers to assess the severity of incompatible changes, the probability of encountering them, and can perhaps even provide information about the detectability of changes, such as the expected recall, precision and performance.

\paragraph{Data provenance.}
Component upgrades are a collaboration between upstream developers and downstream developers. Communication is key to any successful collaboration.
Semantic version numbers are a quickly-understood
shorthand for communicating about the impacts of changes, and suffice
in many contexts, particularly low-consequence ones. However, as we
have established throughout this essay, the general version
compatibility problem is complicated. Below, we sketch a path forward,
which aims to provide both version numbers (when they suffice) backed up by
more detailed information (when needed by developers).

We advocate for the use of \emph{data provenance}, as defined in \cite{buneman2001and}: \textit{``where a piece of data came from and the process by which it arrived in the database''}. Currently, in semantic versioning, upstream developers provide a version number, a set of code changes, and perhaps a textual changelog. We envision a future where upstream developers provide, along with the version number, metadata encoding the reasoning behind the choice of that number, as supported by output from a versioning calculator (possibly stored in a database). In this future, downstream developers can use the provenance information to (1) establish trust in a versioning scheme, and (2) help them to assess where additional verification checks for a particular client should be performed. Such metadata could also include developer identities as well as richer reputational information that could help impede dependency attacks like the event-stream incident mentioned in Section~\ref{sec:upstream}.

For instance, an upstream developer making a release would provide the new version of the changed component, but would also attach a (likely machine-readable) record of the steps that were performed to calculate this version, such as ``\textit{Revapi checks and cross version testing were performed and revealed no API-breaking changes}''.  A downstream developer using this component would then be in a position to make an informed decision based on this information whether to (1) upgrade without further checks, (2) upgrade with additional checks, (3) not upgrade or (4) always automatically upgrade that component from now on (deciding that the processes used upstream are sufficient to ensure compatibility, or more pragmatically, to ensure that the cost of compatibility-breaking changes is sufficiently low). For option (4), the downstream developer would use version ranges or similar syntactic constructs to enable the automated upgrade feature of the respective package manager.

\paragraph{Implementing a calculator.} Figure \ref{fig:calculator} depicts the use of a semantic version calculator from the point of view of the upstream developers. The downstream developer could read off the impacts of the change and decide which of the above tactics to employ.

The versioning calculator could be implemented using a Datalog-based implementation (or some other reasoning tool) which combines the output of a range of program analysis tools. A simple Datalog-based presentation of those checks looks particularly promising for a number of reasons: 
\begin{enumerate}
	\item Community-specific versioning policies can be written using simple developer-friendly rules.
	\item In a Datalog-based presentation, a taxonomy can be easily represented as predicates. Facts for those predicates are then produced by various program analyses.
	\item Datalog has built-in provenance that can reproduce the reasoning process with the rules that were applied to compute a result (a version number).
        \item Datalog scales well, and has a sound formal foundation with its fixpoint-based semantics.
	\item Datalog is already widely used by the program analysis community.
\end{enumerate}
We anticipate that the calculator implementation itself would be just a shallow layer that combines the results of sophisticated analyses, classifies them within an appropriate taxonomy, and recommends appropriate actions to the developers.


        \section{Conclusion}

The management of dependencies and their life cycle  is an increasingly important aspect of modern
software engineering.  Not getting it right exposes applications to bugs and vulnerabilities. 
We have discussed the challenges that developers face when confronted with published
component upgrades or when themselves publishing upgrades. In brief,
downstream developers must balance the cost of updating their code to
newer published versions against the cost of vulnerabilities and bugs
in older versions. Upstream developers, on the other hand, are constrained to
not unnecessarily break their clients, which can prevent them from
making beneficial changes to their libraries.

Choosing an upgrade strategy for one's software requires a deep
understanding of the nature and impact of upstream and downstream
changes, which requires sophisticated reasoning. And yet, tools for enforcing versioning protocols remain
relatively unsophisticated.
Reasoning about upgrades can be a rich source of sharply-defined problems for program analysis to handle.



We observe that upgrades can be viewed as changes to contracts (from
heavyweight formal contracts down to lightweight type annotations) and
to implementations of contracts, both areas that have been extensively
studied. The lack of explicit contracts on most legacy code remains a
significant obstacle to formal reasoning about such code; research
into contract inference will help with the upgrade problem along with
many others.  Likewise, late binding represents another hurdle, as code
is rarely run with the exact version of a library it was compiled
against.  Instead, efforts are required to embed contract information
within binaries which can be checked at (dynamic) link time.
Additionally, non-functional requirements are equally important and
must be considered when upgrading components as well.  The question of
which APIs are relevant to upgrades is itself both a programming
language design and program analysis problem.

Several recent developments lead us to believe that the time is
ripe for serious investment in the development of tools for version
checking:

\begin{itemize}
\item Widely-used automated builds create opportunities to deploy
  dynamic analysis, for instance by monitoring test executions via
  instrumentation or sampling. This infrastructure can be exploited to
  devise hybrid analyses which have shown to be effective in
  addressing recall issues with pure static analysis.
  
\item Recent advances in test and input generation can be exploited to
  further improve the recall of dynamic analyses, and to detect
  faults.

\item Analyzing the impact of changes is basically an incremental
  static analysis problem. Effective incremental analysis can address
  scalability. This has been demonstrated recently by
  deploying separation-logic-based static analysis on a large scale at
  Facebook in the context of the \textit{Infer} tool. There is active research
  on the incremental computation of fixpoints in Datalog, which will
  have a direct impact on Datalog-based static analyses in the near
  future.

\item There are new models of deploying program analysis as a service
  without having to proselytize developer communities. With consent
  but without requiring buy-in, academic and industrial third parties
  currently automatically create pull requests on GitHub from analysis
  results. The third-party approach creates novel opportunities to
  validate program analyses.
  
\end{itemize}

Many prerequisites for better versioning calculators exist today,
both in terms of tools, as discussed above, and data, in the form of
labelled artifact repositories like Maven's. We believe that integrating
the tools and the data can lead to powerful new semantic versioning calculators
which will help both upstream and downstream developers.

\paragraph{Limitations of semantic versioning.} 
Semantic versioning collapses the universe of possible changes in a component into three integers and asks the upstream developer to estimate the consequences of their changes

A taxonomy like the one described in Section~\ref{sec:taxonomy} makes explicit
the many axes along which changes can occur. At an individual change
level, some changes are relevant to some clients but not others. The
same is true at an axis level: some classes of clients are indifferent
to entire axes of changes---for instance, prototype code might not
care about performance.

In a sense, we wish to re-imagine the semantic versioning manifesto in
a more pragmatic light.  The {\em Power of 10} rules for developing
safety-critical software offer inspiration here~\cite{Holz06}.
Providing safe coding guidelines is a somehow limitless and subjective
endeavour.  And, like the semantic versioning manifesto, safe coding guidelines often
rely on programmers to self-check their conformance.  Holzmann
believed that such guidelines ``offer limited benefit'', instead arguing:

\begin{quote}
  ``{\em To be effective, though, the set of rules must be small, and
    it must be clear enough that users can easily understand and
    remember it. In addition, the rules must be specific enough that
    users can check them thoroughly and mechanically.}''
\end{quote}

His emphasis was on rules that could be mechanically enforced, rather
than rules that relied on the whims of programmers for adherence.  In
these respects, we agree---tool support is essential and, without
this, it is difficult for the idea of semantic versioning to realize
its potential.

We argue that the semantic versioning manifesto as stated is
woefully idealistic, and that a more pragmatic manifesto is
needed---one which pays particular attention to machine checkable
rules. As it stands, we rely on the best efforts of developers
generally without tools.  An industrial colleague commented that
``semver is a lie not worth the business risk'', at least in the
Python and JS ecosystems.

\paragraph{A way forward.} Our belief is that a more pragmatic manifesto, along with tools that provide more certainty,
 would be valuable for developers.
However, it is possible that work on better versioning might reveal
that the semantic versioning manifesto is fundamentally broken and must be replaced with
a new, multidimensional paradigm. Upgrade compatibility may perhaps be too complicated to represent by
a triple of numbers.

To that end, taxonomies enable communities and individual downstream developers to look beyond
the three numbers of a semantic version and to make fine-grained
decisions on the compatibility of new upstream components.
Managing evolution is not a pure technical problem but has social
aspects and is embedded in a larger context.
Any versioning solution that seeks to have impact must be appropriate for its context.

Our generic taxonomy presented many axes to consider when deciding when
to upgrade, including the presence of bug fixes and new features, changes in non-functional properties, and
various degrees of breaking changes.
An ecosystem-appropriate taxonomy of compatibility-related issues formal enough to reason about
but that is still practical would be an important step towards semantic semantic versioning.
Existing and novel analyses have a role to play in producing information
that can be incorporated into taxonomies and leveraged by communities to build tools that suit their needs.

Using a broadly-accepted taxonomy, semantic version calculators could be declaratively composed, offsetting each others' shortcomings.  Developing declarative rule languages that operate on a taxonomy is not technically challenging---generic formats and tools exist and could be used. Once again, the challenge is a social one, i.e. to get researchers, tool builders and developers to agree on formats.

The ultimate validation would be to create tools that are accepted and used in practice. These tools must result in reduced technical lag (and therefore fewer propagated bugs and vulnerabilities), fewer compatibility errors, or both. 

In this essay, we have outlined the space of tools to understand
versioning and examined some of the important axes of rotation.
Today's developer ecosystem---with publicly-available releases,
version repositories, continuous integration, and issue
trackers---provide fertile ground for this research.  We hope this
essay serves as a "call to arms" for researchers with a broad
range of interests, including:
\begin{itemize}
\item formal and lightweight contracts: detecting and verifying them, both
  statically and dynamically, and including incremental and high-recall analysis;
\item API surfaces: approaches from language design through to hybrid and static analyses to
  delineate the API surface that needs to be reasoned about;
\item taxonomies of component changes and
  tools to reason about them; and,
\item human factors research investigating how all of these above tools can
  be developed in cooperation with specific developer ecosystems.
\end{itemize}
We strongly encourage further development of tools specifically
aimed at version checking.  In our view, the time is right for the
program analysis community to create tools that can have a significant
impact on industry practice.  We invite the community to join us in
working on this important problem.

\hyphenation{Mey-ero-vich}
\paragraph{Acknowledgements.} 
We thank Chintan Patel for developing tools to identify some of the breaking changes we used as examples, and Laurian Angelescu, Max Dietrich, Leo Meyerovich, and Lucas Wojciechowski for valuable insights from an industry perspective.

	\bibliography{references,abbrevs-full,djp-references}

	
	
\end{document}